\documentclass[]{aastex631}
\shorttitle{A BL-type solar dynamo operating in the bulk of the convection zone}
\shortauthors{Zhang and Jiang}
\graphicspath{{./}{figures/}}
\begin{document}

\title{A Babcock-Leighton-type Solar Dynamo Operating in the Bulk of the Convection Zone}

\correspondingauthor{Jie Jiang}
\email{jiejiang@buaa.edu.cn}

\author[0000-0002-2219-9829]{Zebin Zhang}
\affiliation{School of Space and Environment, Beihang University, Beijing, People's Republic of China}

\author[0000-0001-5002-0577]{Jie Jiang}
\affiliation{School of Space and Environment, Beihang University, Beijing, People's Republic of China}
\affiliation{Key Laboratory of Space Environment monitoring and Information Processing of MIIT, Beijing, China}



\begin{abstract}
The toroidal magnetic field is assumed to be generated in the
tachocline in most Babcock-Leighton (BL)-type solar
dynamo models, in which the poloidal field is produced
by the emergence and subsequent dispersal of sunspot
groups. However, magnetic activity of fully convective
stars and MHD simulations of global stellar convection
have recently raised serious doubts regarding the
importance of the tachocline in the generation of
the toroidal field. In this study, we aim to develop a
new BL-type dynamo model, in which the dynamo operates
mainly within the bulk of the convection zone. Our 2D
model includes the effect of solar-like differential
rotation, one-cell meridional flow, near-surface radial
pumping, strong turbulent diffusion, BL-type poloidal
source, and nonlinear back-reaction of the magnetic
field on its source with a vertical outer boundary
condition. The model leads to a simple dipolar
configuration of the poloidal field that has the
dominant latitudinal component, which is wound up by
the latitudinal shear within the bulk of the convection
zone to generate the toroidal flux. As a result, the tachocline
plays a negligible role in the model.
The model reproduces the basic properties of the solar
cycle, including (a) approximately 11 yr cycle period
and 18 yr extended cycle period; (b) equatorward
propagation of the antisymmetric toroidal field
starting from high latitudes; and (c) polar field
evolution that is consistent with observations. Our model opens the possibility for
a paradigm shift in understanding the solar cycle to transition from the classical
flux transport dynamo.
\end{abstract}



\section{Introduction}
The dynamo process is responsible for
the solar magnetic cycle \citep{Hathaway2015}. The core
concept for the dynamo is that the poloidal and toroidal
fields sustain each other to form periodic large-scale
magnetic activity \citep{Karak2014, Charbonneau2020}.
Differential rotation stretches the poloidal field to
generate the toroidal field ($\Omega$-effect). There
are several controversial mechanisms for the poloidal field
regeneration ($\alpha$-effect), and the Babcock-Leighton (BL)
mechanism is one of them. The essential idea behind the
mechanism is that the poloidal field is generated by 
the observed behaviors of sunspot
emergence and the subsequent evolution of the magnetic
field at the surface.

The BL mechanism has received significant observational
support during the last decade. In this mechanism, the sunspot 
group tilt angles play a key role in the construction of the 
poloidal source term. \cite{Espuig2010}, \cite{Kitchatinov2011}, and \cite{Jiao2021} reveal
that the observed sunspot group tilt angles vary
systematically from cycle to cycle. The weak cycle 24 might
result from abnormal tilt angles
of sunspot groups in cycle 23 \citep{Jiang2015}. 
Furthermore, observations confirm the correlation between the 
polar field strength at cycle minimum and the next cycle strength 
implied by the BL mechanism \citep{Schatten1978,Jaramillo2013, Jiang2018}.

In Babcock's original scenario \citep{Babcock1961}, the
latitudinal differential rotation stretches the poloidal magnetic
field that is represented by the global dipole to generate the toroidal field.
\cite{Cameron2015} find that the latitudinal differential rotation 
is by far the dominant generator of the net toroidal flux.
The important role of the latitudinal shear in the
toroidal field generation is emphasized by previous
studies, such as \cite{Guerrero2007} and \cite{Jaramillo2009}.
However, there appears to be a long-standing consensus that the toroidal field 
is generated by the radial differential rotation in the tachocline; that is,
the narrow radial shear layer at the bottom of the convection 
zone \citep{Spiegel1992}.

Since their development in the 1990s, flux transport dynamo (FTD) models have emerged as a popular 
paradigm for the understanding of the solar cycle \citep{Wang1991, Choudhuri1995, Durney1995}.
The BL mechanism is usually incorporated in these models,
which are characterized by an important
role played by the meridional flow. Moreover, the toroidal field 
is created in the tachocline and rises owing to magnetic 
buoyancy to form sunspots. The scenario for the toroidal 
field's generation and emergence is supported by thin flux tube
simulations \citep[for reviews, see][]{Fan2009}, which
can produce the proper
latitudinal dependence of tilt angles with an initial
field as strong as 10$^5$ G \citep{D'Silva1993, Caligari1995}.
The tachocline's significant radial shear and subadiabatic
stratification may allow the magnetic fields to be
amplified so tremendously before being subjected to magnetic
buoyancy instabilities. These are arguments in favor of
the tachocline for the toroidal field location.

However, the last few decades have seen problems with FTD models,
which are directly
caused by the effect of the tachocline, despite their
effectiveness in producing some solar cycle features.
In the polar latitudes, the tachocline has a strong radial shear,
which causes the toroidal flux to prevail in polar
regions \citep{Dikpati1999, Nandy2002, Guerrero2004, Lemerle2017}.
The concentrated toroidal flux at high latitudes causes a
rather weak ratio between the toroidal field at the
activity belt and polar fields \citep{Kitchatinov2012}.
There are several other problems with FTD models that are produced
indirectly by the tachocline's effect. For example, FTD
models tend to generate the quadrupole-like
parity \citep{Dikpati2001, Chatterjee2004, Hotta2010} and
are sensitive to the profile of the deep meridional
flow, which is still controversial \citep[for reviews, see ][]{Choudhuri2021}.

Recent MHD simulations of global solar/stellar convection
provide a new scenario for 
the generation of the toroidal flux. 
\cite{Nelson2013, Nelson2014} and \cite{Chen2021} show that
convective simulations without the tachocline spontaneously
generate rope-like structures of magnetic flux that rise
to the top of the simulation domain in a solar-like
manner. \cite{Fan2014}, \cite{Guerrero2016},
and \cite{Kapyla2017}
demonstrate that solar-like large-scale magnetic fields
can be produced entirely within a convection zone without extending the
simulation down to the tachocline.
\cite{Kapyla2021} shows the similarity of large-scale
field generation between partially and fully convective
stars. Progress in MHD simulations provides an impetus
to rethink the roles of the tachocline in the BL-type
dynamo.

Theoretically, \cite{Spruit2011} argues that a tachocline
that has a stable stratification does not provide significant
shear stresses, with which a field could be amplified.
\cite{Kitchatinov2017b} argue that the
radial field should be small near the base of the convection
zone. The tachocline is unimportant
for the dynamo, even if there is a strong radial gradient of
rotation in the tachocline. See \cite{Brandenburg2005} for
other arguments against the tachocline for the toroidal
field origination.

Another persuasive question on the importance of the
tachocline is proposed by stellar magnetic activity.
If a tachocline was really crucial for a dynamo to work,
then one might expect a break in magnetic activity toward
fully convective late-M dwarfs \citep{Charbonneau2016}.
However, it appears to be continuous across
the transition to full convection for most activity indicators, 
such as flare activity \citep{Lin2019, Yang2019}. 
In particular, the X-ray activity-rotation relationship of 
fully convective M-type stars is in line with partially 
convective stars \citep{Wright2016}.

All of the evidence that is listed here questions the widely accepted
opinion that the toroidal field is generated by the radial shear
in the tachocline. In this regard, a dynamo model
operating in the bulk of the convection zone in the framework
of the BL mechanism is required. \cite{Cameron2017a} have taken the first step toward the
new generation, in the sense that
the tachocline does not play an essential role anymore, of the BL-type dynamo.
They assume that the toroidal flux is generated by the
radial shear in the near-surface shear layer (NSSL)
and by the latitudinal shear in the bulk of the
convection zone. The
solar-like solutions are obtained from their quasi-1D
model. However, the model cannot elaborate the radial distributions and
configurations of the toroidal and poloidal fluxes, which 
requires the development of a 2D BL-type model.

The purpose of the present paper is to develop a 2D
BL-type dynamo model, in which the toroidal field is
generated in the bulk of the convection zone. The model
is consistent with the key property of the standard
Babcock's scenario, which is that the latitudinal
differential rotation stretches the simple dipole to
generate the toroidal field.

This paper is organized as follows. The new BL-type dynamo
model is described in Section \ref{sec:model}. A reference
model is presented in detail in Section \ref{sec:referenceSolution}.
We demonstrate that our model is independent of the tachocline
in Section \ref{sec:negligibleRoleTachocline}. Comparisons with two
FTD models are presented in Section \ref{sec:ComparisonsModels}.
Finally, we summarize and discuss our results in Section \ref{sec:con}.

\section{MODEL} \label{sec:model}
The axisymmetric large-scale magnetic field is 
expressed in spherical coordinates as
\begin{equation}
  \textbf{B}(r,\theta,t)=B(r,\theta,t) \hat{\textbf{e}}_\phi+
  \nabla\times
  \left[A(r,\theta,t)\hat{\textbf{e}}_\phi\right],\label{eq1}
\end{equation}
where $B(r,\theta,t)\hat{\textbf{e}}_\phi$ represents
the toroidal field and the curl of a magnetic vector potential in
$\phi$ direction represents the poloidal field.
In the kinematic framework, the large-scale flow fields are prescribed
as
\begin{equation}
  \textbf{u}(r,\theta)=r \sin\theta \Omega(r,\theta) \hat
  {\textbf{e}}_\phi+\textbf{u}_{p}(r,\theta),\label{eq2}
 \end{equation}
where $\textbf{u}_{p}(r,\theta)$ represents the meridional
flow, and $\Omega$ is the angular velocity. The BL-type $\alpha \Omega$ dynamo equations
\citep[for reviews, see ][]{Charbonneau2020} are
\begin{equation}
  \frac{\partial A}{\partial t}+\frac{1}{s}(\textbf{u}_{p}\cdot\nabla)(sA)
  =\eta\left(\nabla^{2}-\frac{1}{s^{2}}\right)A+S_{BL},\label{eq3}
\end{equation}

\begin{equation}
  \frac{\partial B}{\partial t}+\frac{1}{r}\left[\frac{\partial(ru_{r}B)}
  {\partial r}+\frac{\partial(u_{\theta}B)}{\partial\theta}\right]=\eta\left(\nabla^{2}-\frac{1}
  {s^2}\right)B+s(\textbf{B}_{p}\cdot\nabla\Omega)+\frac{1}{r}\frac{d\eta}
  {dr}\frac{\partial(rB)}{\partial r},\label{eq4}
\end{equation}
where $s=r\sin\theta$, and $\eta$ represents the turbulent
diffusivity. $S_{BL}$ is the BL-type source term that describes the regeneration of the poloidal field
at the solar surface. The second term on the right-hand of
Equation~(\ref{eq4}) is the source term of the toroidal field,
which is equal to
\begin{equation}
  s(\textbf{B}_{r}\cdot
   \nabla_r\Omega)+s(\textbf{B}_{\theta}\cdot \nabla_
   \theta\Omega),\label{eq5}
\end{equation}
where the former term means that $B_r$ is stretched by the
radial shear, and the latter term represents that
$B_\theta$ is stretched by the latitudinal shear.

We specify the profiles of $\textbf{u}_{p}$,
$\Omega$, $\eta$, and $S_{BL}$ in next subsections.
The computational domain of our model is 0.65$R_\odot
\leq r \leq $ $R_\odot, 0 \leq \theta \leq \pi$. The outer
boundary condition is that the field is vertical based on
the constraint by \cite{Cameron2012}. Accordingly, we use
$\partial (rA)/ \partial r = 0$, $B=0$ at
$r = R_\odot$. The bottom boundary matches a perfect
conductor, which means that $A = 0$,
$\partial(rB) / \partial r = 0$ at $r = 0.65R_\odot$.
At poles, $A = B = 0$. Our model is computed using the
code SURYA developed by A.R. Choudhuri and his
colleagues \citep{Dikpati1994,Chatterjee2004}.

\subsection{BL-type source term}
We define $S_{BL}$ in Equation~(\ref{eq3}) as
\begin{equation}
  S_{BL}(r,\theta,t)=\frac{\alpha(r,\theta) \bar{B}(\theta,t)}
  {1+[|\bar{B}(\theta,t)|/B_{0}]^{1/4}},\label{eq6}
\end{equation}
where
\begin{equation}
  \alpha(r,\theta)=\frac{\alpha_{0}f(\theta)}{2}\left[1+\rm
  erf\left(\frac{r-0.95R_\odot}{0.01R_\odot}\right)\right].\label{eq7}
\end{equation}
In the BL mechanism, the regeneration process of the
poloidal field appears at the solar surface, due to the decay
of tilted sunspots. Hence, the source term $S_{BL}$ represented here
is confined above 0.95$R_\odot$. This could produce a near-surface
dipole moment, which is proportional to the mean (area normalized)
toroidal field over the whole convection zone,
\begin{equation}
  \bar{B}(\theta,t)=\int_{0.7R_\odot}^{R_\odot}B(r,\theta,t)
    rdr/\int_{0.7R_\odot}^{R_\odot}rdr.\label{eq8}
\end{equation}
In contrast, $\bar{B}(r,\theta,t)$ is usually defined as the
toroidal field in the tachocline in FTD models
\citep[e.g.,][]{Dikpati1999, Chatterjee2004}.

The latitude dependence of the source term $S_{BL}$ is
\begin{equation}
  f(\theta)=\cos\theta \sin^n\theta,\label{eq9}
\end{equation}
where $\cos\theta$ reflects the latitude dependence of the tilt
angles caused by the Coriolis force. 
% \st{In our model,
% we set} $n = 1$ \st{to ensure that the emergence probability per unit length
% of the toroidal field lines is constant in different latitudes.
% Our model contains no artificial constraints on the
% flux emergence, while latitudes of the
% flux emergence are forced to low latitudes by
% setting the} $n$-value \st{larger than 1 in some previous models}
% \citep[e.g.,][]{Jaramillo2011, Karak2014b, MIESCH20161571}.
Parameter $n$ controls the emergence latitude
of the toroidal field. Because the buoyancy process of the 
toroidal field is still an open question, 
we treat $n$ as a free parameter, which is specified 
based on the condition that the results of our model fit the 
observations.

Generally, in a nonlinear dynamo model, a limit 
cycle (periodic solution) emerges from a supercritical 
Hopf bifurcation when the dynamo number exceeds its 
critical value \citep{Tobias1995, Cameron2017b}.
As an amplitude limit mechanism, 
the nonlinear algebraic $\alpha$ quenching 
term $\alpha /[1+(\bar{B}/B_{0})^2]$ is widely
used. This constrains the magnetic fields not to exceed 
the equipartition field $B_{0}$. However, when we 
consider this term in our model, the Hopf bifurcation 
tends to be subcritical, in which condition the 
saturation magnetic fields will largely exceed $B_{0}$.
To achieve our desired quenching effect,
we find that the quenching form needs to be 
$\alpha /[1+(|\bar{B}|/B_{0})^{1/4}]$.
We chose the equipartition field 10$^4$ G for $B_0$, and the units of 
magnetic fields in all solutions are the same as $B_0$. Besides 
the pure BL $\alpha$-effect concentrated near the solar surface 
presented above, we do not include any other $\alpha$-terms in our model.

\subsection{Differential rotation}
We use an analytical profile of the angular velocity close 
to helioseismic results \citep{Schou1998}. The profile is 
given by
\begin{equation}
  \Omega(r,\theta) = \Omega_{RZ} + (\Omega_{CZ}-\Omega_{RZ})
  \Omega_1,\label{eq10}
\end{equation}
\begin{equation}
  \Omega_1=\frac{1}{2}\left[1+\rm erf\left(\frac{r-0.7R_\odot}{0.04R_\odot}\right)\right],\label{eq11}
\end{equation}
where $\Omega_{RZ}/{2\pi}=432.8$ nHz, $\Omega_{CZ}/2
\pi=470.308-62.79\cos^{2}(\theta)-67.87\cos^{4}(\theta)-
16.808(r-0.95$$R_\odot)/0.05$R$_\odot$ nHz for $r>0.95$$R_\odot$, and $\Omega_{CZ}/2\pi=470.308-62.79
\cos^{2}(\theta)-67.87\cos^{4}(\theta)$ nHz for $r\leq 0.95$$R_\odot$.

%figure 1
\begin{figure}[h!]
  \plotone{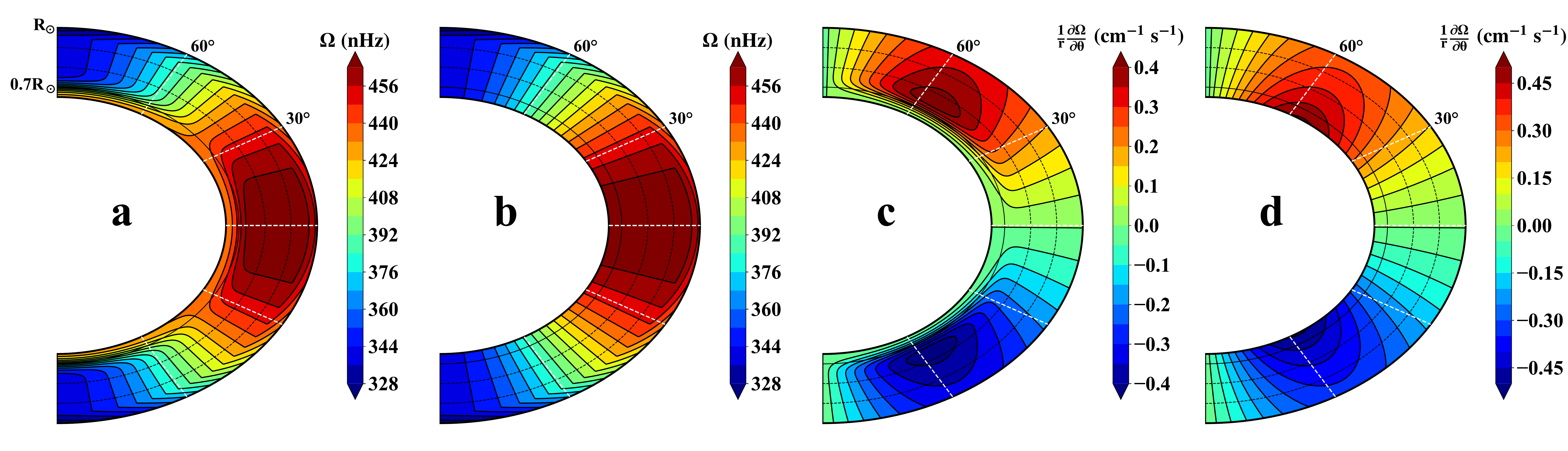}
  \caption{Contours of angular velocity profiles. (a) Angular velocity $\Omega$ of the reference model. (b) Angular velocity $\Omega$ with artificially shutting of the radial shear in the tachocline used in Section 3.2.2. (c) Latitudinal gradient of the angular velocity presented in (a). (d) Latitudinal gradient of the angular velocity presented in (b).}
  \label{fig:diff_rot}
  \end{figure}

The distribution of $\Omega(r,\theta)$ generated by the above expression is shown in
Figure \ref{fig:diff_rot}(a). The radial shear concentrates in the tachocline and the NSSL, while the latitudinal shear concentrates in the bulk of the convection zone.
To evaluate the role of the radial shear in the tachocline in Section \ref{sec:negligibleRoleTachocline},
we set $\Omega_1 = 1$ to obtain the profile that is shown in Figure \ref{fig:diff_rot}(b), which 
contains no radial shear in the tachocline.
Figures \ref{fig:diff_rot}(c) and (d) show the distributions of
$(1/r)(\partial\Omega/\partial\theta)$
for the rotation profiles shown in Figures \ref{fig:diff_rot}(a) and (b), respectively.
They show the strong latitudinal shear in the bulk of the convection zone.
The maximum value of $(1/r)(\partial\Omega/\partial\theta)$
appears at about 55$^\circ$ latitude and the value of $(1/r)(\partial\Omega/\partial\theta)$
decreases with depth. Note that when the radial shear 
is removed, the latitudinal shear is stronger than that 
with the radial shear in the tachocline, especially at 
the base of the convection zone.

\subsection{Meridional flow}
We define the stream function $\psi$ that satisfies the equation,
 \begin{equation}
    \psi r \sin\theta = \psi_0(r-R_p)\sin\left[\frac{\pi(r-R_p)}{R_\odot-R_p}\right](1-e^{-\beta_1\theta^\varepsilon })[1-e^{\beta_2(\theta-\pi/2)}]e^{-[(r-r_0)/\Gamma]^2 }\sin\theta,\label{eq12}
\end{equation}
where $\beta_1 = 1.5, \beta_2 = 1.3, \varepsilon = 2.0000001, r_0 = (R_\odot-R_b)/3.5, \Gamma =3.47 \times 10^8 m$ 
and $R_p = 0.7R_\odot$. According to 
$\rho \textbf{u}_p = \nabla\times[\psi(r,\theta)] \hat{\textbf{e}}_\phi$, 
where $\rho = C(R_\odot/r-0.95)^{3/2}$, the profile of the meridional flow
$ \textbf{u}_p$ could be derived.
By setting $\psi_0/C$ = 33.4 m s$^{-1}$, the amplitude of the 
surface velocity reaches a maximum of 20 m s$^{-1}$ at 45$^\circ$.  
The inner return flow decreases to zero at $r$ = 0.7$R_\odot$. 
This profile is similar to Equation (5) of \cite{Karak2016}, 
except that we add a term $\sin \theta$ in Equation~(\ref{eq12}) 
to decrease the flow speed near the poles. 

The typical amplitudes of 10 - 20 m s$^{-1}$ for 
the surface poleward flow have been confirmed by several 
measurements \citep{Hathaway2010, Ulrich2010, Basu2010} 
but there is no consensus on the inner flow 
\citep{Zhao2013,Schad2013,Rajaguru2015,Gizon2020}.
The meridional flow used in our model is similar to the 
helioseismic results of \cite{Jackiewicz2015} in terms of 
the return flow starting in shallower layers.

\begin{figure}[t!]
  \plotone{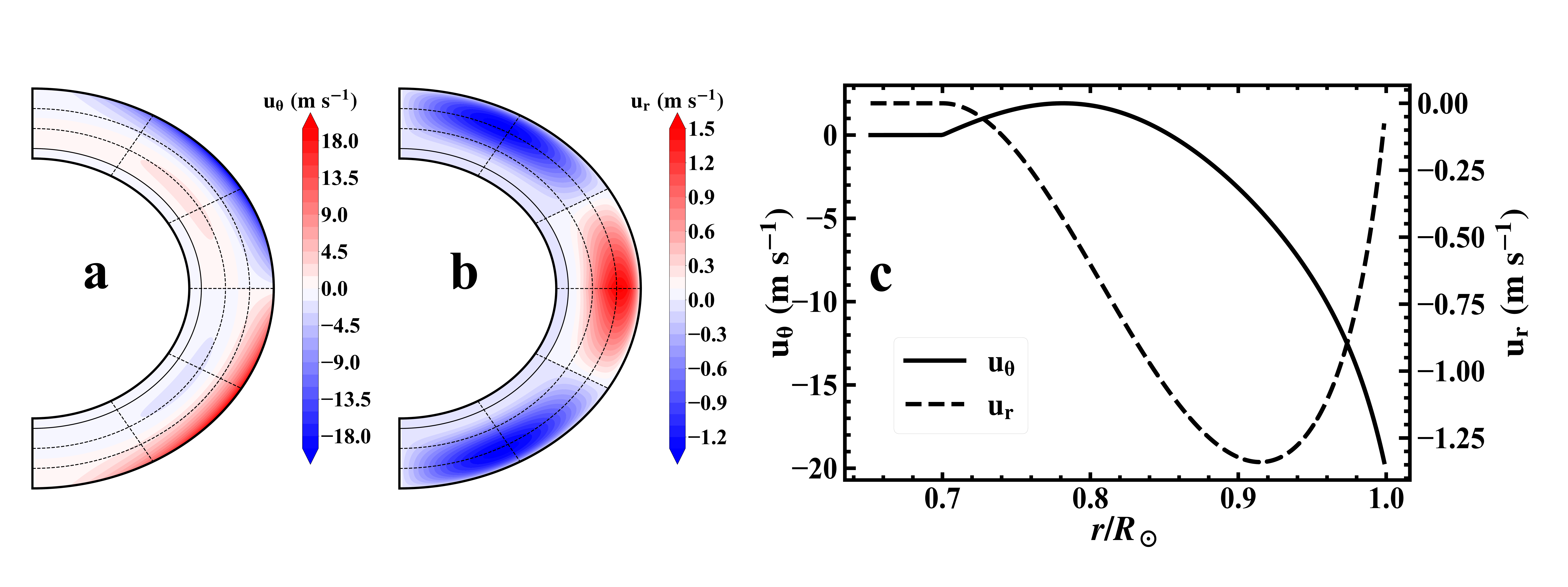}
  \caption{Meridional flow of the reference model. (a) Latitudinal component, $u_\theta$. (b) Radial component, $u_r$. (c) Radial variations of $u_\theta$ at 45$^\circ$ latitude (solid line) and $u_r$ at 60$^\circ$ latitude (dashed line).}
   \label{fig:MC}
\end{figure}

The distributions of $u_\theta$ and $u_r$ in the 
meridional plane are shown in Figures \ref{fig:MC}(a) and (b), 
respectively. Figure \ref{fig:MC}(c) shows the radial 
variations of $u_\theta$ at 45$^\circ$ latitudes and $u_r$ 
at 60$^\circ$ latitudes. The return flow starts from 0.86$R_\odot$ 
and the average value of $u_\theta R_\odot/r$ is 1.6 m s$^{-1}$
in the range 0.7-0.86$R_\odot$ where the toroidal
flux resides.

%figure 2

\subsection{Turbulent magnetic pumping}
The pumping effect is caused by the inhomogeneity of
turbulence, which contains density (density pumping),
turbulent velocity (turbulent pumping), convection
structure (topological pumping), and turbulent
diffusivity (diamagnetic pumping).
\cite{Brandenburg1992} and \cite{Kitchatinov2012}
include the diamagnetic pumping at the
bottom of the convection zone in their dynamo models.
For the first time, \cite{Guerrero2008} introduce the 
latitudinal and radial pumping throughout the whole 
convection zone in a BL-type dynamo model. 

In this paper, we only consider the near-surface pumping to ensure
the surface field evolution from the BL-type dynamo model
consistent with that from the surface flux 
transport (SFT) model, which can well describe the 
large-scale field evolution over the solar 
surface \citep{Wang1989, Baumann2004, Mackay2012, Jiang2014, Yeates2015, Petrovay2019, Wang2020}. 
The effect of near-surface pumping on the surface poloidal 
field evolution in a BL-type dynamo is first proposed 
by \cite{Cameron2012}. 
\cite{Jiang2013}, \cite{Karak2016}, and \cite{Karak2017} further 
demonstrate its importance in BL-type dynamo models.

Downward pumping could be viewed
as an advection of the magnetic field with a certain velocity,
so we replace
$u_{r}(r,\theta)$ as $u_{r}(r,\theta)+\gamma(r)$,
where
\begin{equation}
  \gamma(r)=-\frac{\gamma_{0}}{2}\left[1+\rm erf\left(\frac{r-r_{p}}
  {0.04R_\odot}\right)\right].\label{eq15}
\end{equation}
There are no strict constraints on the total 
strength $\gamma_{0}$ 
or the penetration depth $r_p$ of the pumping. For simplicity, 
the pumping profile does not depend on latitudes.

\subsection{Turbulent magnetic diffusivity}
The value of turbulent magnetic diffusivity can be estimated by the 
mixing-length theory (MLT), which gives the diffusivity 
in order 
of $10^{13}$ cm$^2$ s$^{-1}$ \citep{Jaramillo2011}.
However, the diffusivity used in previous models is 1-4 orders of 
magnitude lower than what the theory suggests 
\citep[see Figure 1 of][]{Jaramillo2011}.
\cite{Karak2016} show that the magnetic diffusivity could be high 
with the introduction of the near-surface pumping because the pumping 
restrains the outward dissipation of magnetic fields.

%figure 3
\begin{figure}[t!]
  \plotone{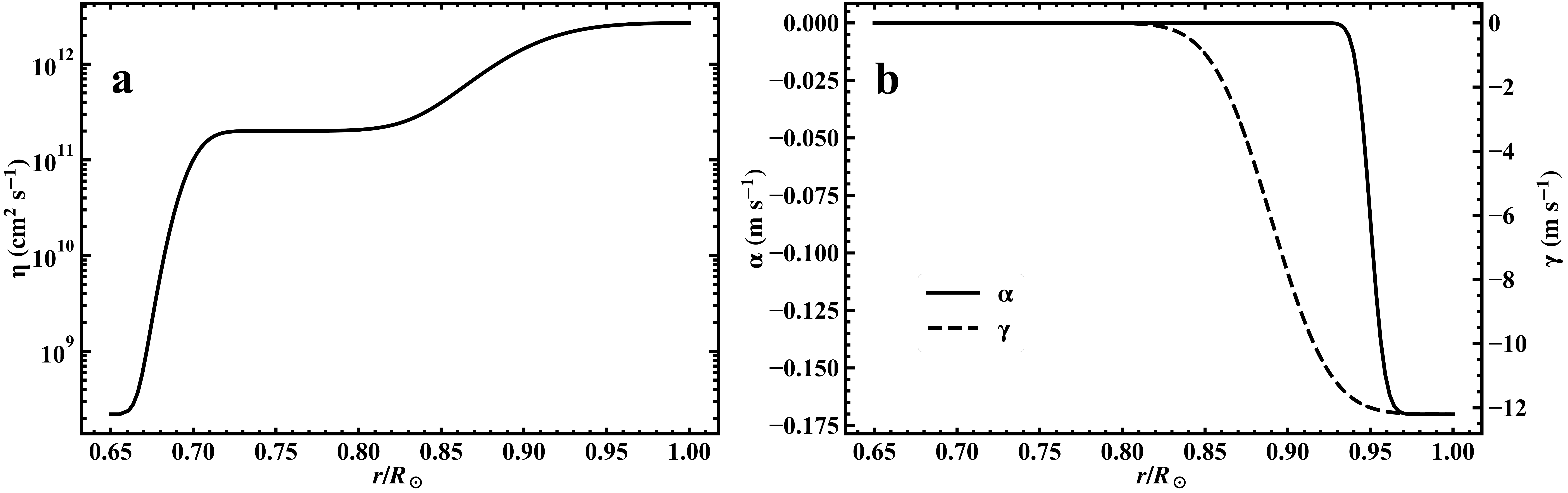}
\caption{Radial variations of (a) turbulent diffusivity and (b) BL-type source term at -45$^\circ$ latitude in  solid line and radial pumping in dashed line, of the reference model.}
\label{fig:parameters}
\end{figure}

We utilize the following diffusivity profile
\begin{equation}
  \eta=\eta_{RZ}+\frac{\eta_{CZ}}{2}\left[1+\rm erf\left(\frac{r-0.7R_\odot}{0.03R_\odot}\right)\right]+\frac{\eta_{S}}{2}\left[1+\rm erf\left(\frac{r-0.9R_\odot}{0.05R_\odot}\right)\right],\label{eq16}
\end{equation}
where 
$\eta_{RZ}=2.2\times10^{8}$ cm$^{2}$ s$^{-1}$,
$\eta_{CZ}=2.0\times10^{11}$ cm$^{2}$ s$^{-1}$, and
$\eta_{S}=2.5\times10^{12}$ cm$^{2}$ s$^{-1}$.
The diffusivity at the surface is represented by $\eta_{S}$, which
is close to observations and estimations based on the MLT.
In the bulk of the convection zone, the diffusivity is
represented by $\eta_{CZ}$, which is slightly lower than that at the surface.
The diffusivity profile is shown in Figure \ref{fig:parameters}(a).

\section{Results}
\subsection{A reference model} \label{sec:referenceSolution}
In this section we present a representative solar-like
solution, calculated with parameter values
% \st{$r_p=0.88R_\odot$, $\gamma_0=35$ m s$^{-1}$, and
% $\alpha_0$ = 0.135 m s$^{-1}$.} 
$r_p=0.89R_\odot$, $\gamma_0=12.2$ m s$^{-1}$, $n=6$. 
The profiles of the pumping and the source term at 45$^\circ$ 
in the southern hemisphere are shown in Figure 
\ref{fig:parameters}(b). 
% \st{The pumping profile is
% consistent with the 1D model used by} \cite{Cameron2017}.
% \st{The value of} $\alpha_0$ \st{is several times weaker than
% that of} \cite{Cameron2017} \st{since the nonlinear effect
% in the form of} Equation~(\ref{eq6}) \st{is included
% in our model}. \st{The solution is independent of the choice
% of initial conditions. After the simulation starts,
% transients associated with initial conditions disappear,
% and the magnetic field grows to saturation. Here we
% employ an equator-symmetric initial condition. The
% poloidal field is set to be zero, whereas the toroidal field}
% $B(r, \theta)$ \st{is set to be} $B(r, \theta) = \cos(2\theta)\sin[\pi(r-0.65$R$_\odot)/0.35R_\odot]$
% \st{for} $r \geq 0.65 R_\odot$. 
For the reference case, the critical value for 
dipolar parity is $\alpha_c$ = 0.8 m s$^{-1}$.
Because the solar dynamo is estimated to be 
about 10$\%$ supercritical \citep{Kitchatinov2017a}, 
we set $\alpha_0$ = 0.9 m s$^{-1}$ in simulation.
The solution is independent of the choice
of initial conditions. After the simulation starts,
transients associated with initial conditions disappear
and the magnetic field grows to saturation.
Here, we employ an initial field of a mixed parity. 
The poloidal field is set to be 
zero, whereas the toroidal field $B(r, \theta)$ is set to 
be $B(r, \theta) = \sin(2\theta)\sin[\pi(r-0.7R_\odot)/0.3R_\odot]
 +  \sin(\theta)\sin[\pi(r-0.7R_\odot)/0.3R_\odot]$
for $r \geq 0.7 R_\odot$.

\begin{figure}[t!]
  \plotone{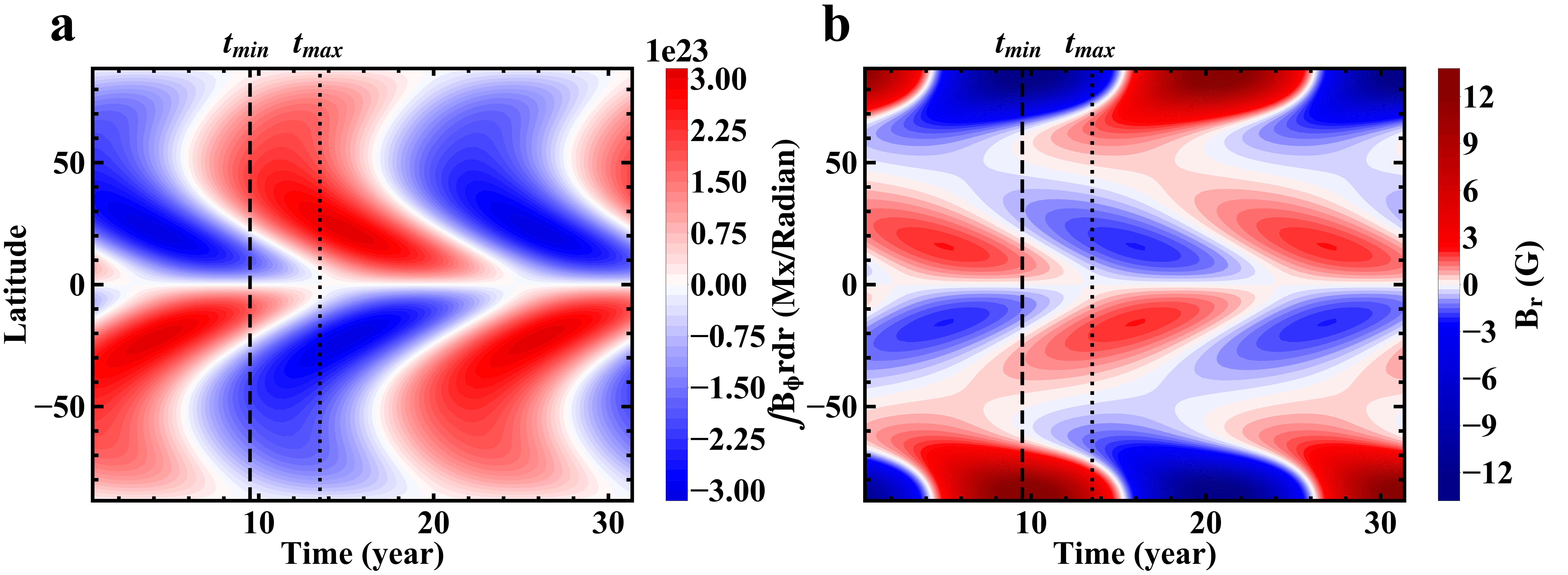}
  \caption{Time-latitude diagrams of (a) radially
  integrated toroidal magnetic flux (per radian) and (b)
  radial field at the surface, for our reference model.
  Cycle minimum ($t_{min}$)
  and cycle maximum ($t_{max}$) are indicated by the vertical
  dashed and dotted lines, respectively.}
  \label{fig:standard1_TorPol}
\end{figure}

\begin{figure}[b!]
  \gridline{\fig{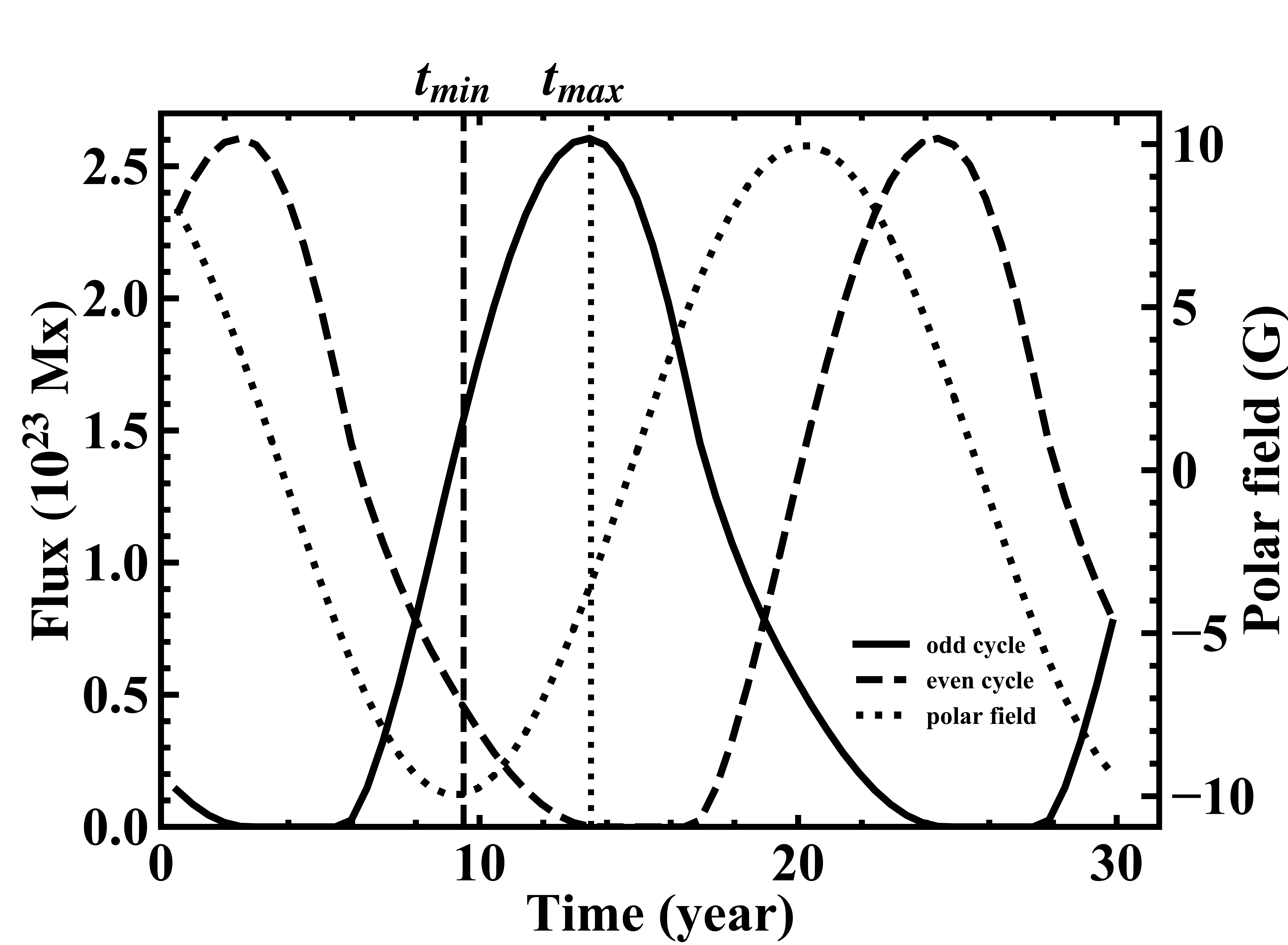}{0.45\textwidth}{}
            }
  \caption{Time evolution of the subsurface toroidal flux and
  polar field, for the reference model.
  The solid and dashed curves represent the unsigned subsurface toroidal flux
  between $0^\circ$ and $40^\circ$ latitudes for odd and even cycles, respectively.
  The dotted line represents the polar field. Cycle minimum ($t_{min}$)
  and cycle maximum ($t_{max}$) are indicated by the vertical dashed and dotted lines, respectively.}
  \label{fig:standard_FluxPolF}
\end{figure}

% \st{The toroidal field is anti-symmetric about the equator 
% although a quadrupolar initial condition is used.} 
Figure \ref{fig:standard1_TorPol}(a) shows a time-latitude 
diagram of the subsurface toroidal flux density calculated by
integrating the toroidal field over the range of
0.7-1.0$R_\odot$. 
The toroidal field is antisymmetric
since the dipolar parity always prevails in our model.
The latitudinal migration pattern of the toroidal field
is different from that of FTD models, which usually show concentrated
flux at high latitudes due to the radial shear in the
polar portion of the tachocline. For each cycle, a weak
toroidal field starts from high latitudes along with
another overlapped branch of the previous cycle near the
equator. With the equatorward migration of the toroidal
field, its flux density increases as the cycle evolutes.
This scenario is consistent with
that of the solar cycle and the extended solar cycle
\citep{Howard1980, Wilson1988, McIntosh2014}. Below
40$^\circ$ latitude, 
% \st{the toroidal flux density is
% larger than} $0.3 \times 10^{22}$ Mx/$\circ$.
the toroidal flux density is
larger than 2.5 $\times 10^{23}$ Mx Radian$^{-1}$.
Considering the spatial scale of a typical active
region to be a few degrees, these fluxes are comparable
with that in active regions. According to
\cite{Cameron2019}, the toroidal field at lower
latitudes corresponds to the expected amount of flux
residing in active regions.
The weaker toroidal field at higher latitudes
corresponds to ephemeral regions, which are small
bipolar magnetic regions that contain a maximum total
flux of the order of $10^{20}$ Mx \citep{Harvey1973}.
In Figure \ref{fig:standard1_TorPol}, the vertical dashed line
represents cycle minimum ($t_{min}$) when the toroidal
flux of the last cycle disappears, and vertical dotted line
represents cycle maximum ($t_{max}$)
when the toroidal flux of the new cycle reaches a maximum.
The interval between $t_{min}$ and $t_{max}$ is 
% \st{3.9 years} 
4 yr, which is
close to the average period of the rising phase for solar
cycles \citep{Hathaway2015, Jiang2018}. The solid and dashed curves of Figure \ref{fig:standard_FluxPolF} show the time
evolution of unsigned toroidal flux,
$\int_{50}^{90^\circ}\int_{0.7R_\odot}^{R_\odot}r|B_\phi|    \,dr   \,d\theta $,
of each hemisphere for odd and even cycles, respectively.
The cycle period is 
% \st{10.6} 
10.96 yr based on the interval between successive maxima of the
toroidal flux. This period is consistent with the average period of observed solar activity
\citep{Hathaway2015}. If the cycle period is
calculated as the interval between the start and end of the toroidal flux of a cycle, then it
is 
% \st{16.4 years} 
18.1 yr. This result is 
consistent with approximately the 18 - 22 yr period of the 
extended cycle \citep{Wilson1988}.
The maximum toroidal flux produced in one hemisphere 
% \st{is about} $2.2\times 10^{23}$ 
is about $2.6\times 10^{23}$ Mx, which matches the 
observed magnetic flux at the solar surface \citep{Cameron2015}.

%figure 4

%figure 5

The time-latitude diagram of $B_r$ at the surface is
presented in Figure \ref{fig:standard1_TorPol}(b). 
There is a poleward branch and an equatorward branch 
that are separated by around $55^\circ$. The poleward branch is 
caused by the poleward flow and turbulent diffusion
at the surface. The flux contained in this branch
reverses the polar field polarity of a previous cycle
and sets up the polar field of a new cycle. The
equatorward branch is mainly caused by the equatorward
propagation of the subsurface toroidal field that is
the source of the poloidal field at the surface based on
Equation~(\ref{eq6}). The maximum polar field is about 
% \textit{16 G} 
13.5 G, which is close to observational constraints
\citep{Svalgaard1978}. 
% \textit{In our model, the toroidal field
% in the bulk of the convection zone amounts to thousands
% of gauss. The ratio of the maximum toroidal field,
% $B_{\phi\max}$, and maximum polar field, $B_{r\max}$,
% is $B_{\phi\max}/B_{r\max}$ = 104,
% which is about twice as large as the values listed in Table 1
% of \cite{Dikpati1999} (except the one having cycle period of 77.94 yr). 
% The inward pumping compresses
% and amplifies the poloidal field in the bulk of the
% convection zone. This compression accounts for the high $B_{\phi\max}/B_{r\max}$ and
% the strong toroidal field of several thousand gauss
% in the bulk of the convection zone.}
In our model, the toroidal field
in the bulk of the convection zone amounts to hundreds
of gauss. The ratio of the maximum toroidal field,
$B_{\phi\max}$, and maximum polar field, $B_{r\max}$,
is $B_{\phi\max}/B_{r\max}$ = 45. The inward pumping 
compresses the poloidal field, which helps to increase
the ratio between the toroidal and poloidal fields. 
The dotted line in Figure \ref{fig:standard_FluxPolF}
represents the mean radial field at the surface over
$70^\circ$ latitudes. The mean radial field reaches
its maximum values about 10 G   
at cycle minimum corresponding to the moment when the
toroidal field from the last cycle disappears. This
means that our model reproduces a correct phase difference between the polar field
and flux emergence at the activity belt. The polarity reversal
cycle of the polar fields is also about 11 yr.

\begin{figure}[ht!]
  \gridline{\fig{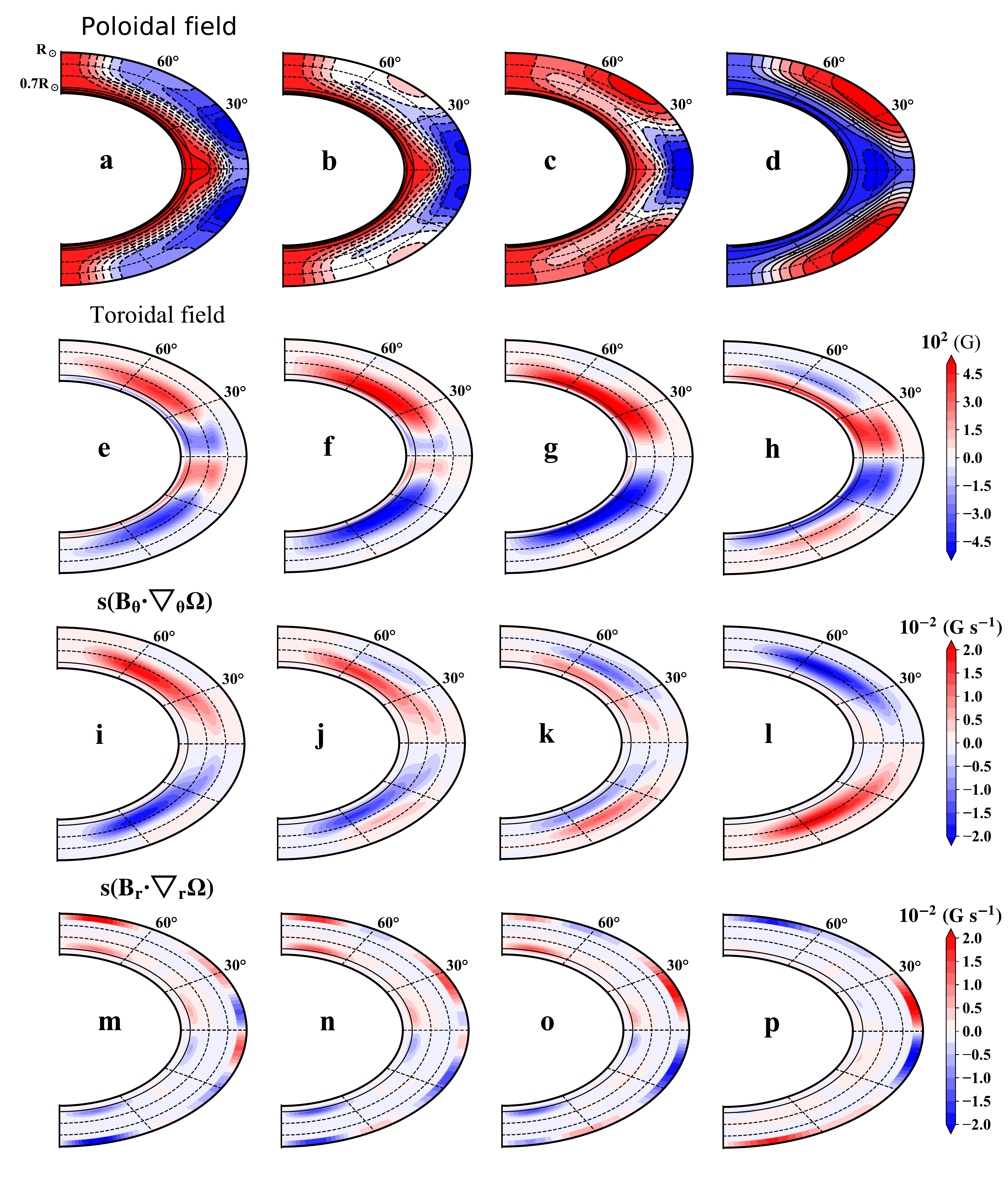}{1.0\textwidth}{}
            }
  \caption{Snapshots of the poloidal field (first row),
  toroidal field (second row), and
  shear source terms (third and fourth rows) over a
  dynamo cycle for the reference model.
  Four columns from left-hand to right-hand correspond to $t_1$ = $t_{min}$,
  $t_2$ = $t_1$ + 1.9 yr, $t_3$ = $t_2$ + 2.1 yr = $t_{max}$, 
  and $t_4$ = $t_3$ + 4.0 yr, respectively.
  Solid (dashed) poloidal field lines are clockwise (counterclockwise).
  }
  \label{fig:stdardCase_snapshots}
\end{figure}

To understand how the model works, we plot the time evolution
of the magnetic fields and the shear source terms for four
successive intervals during a dynamo period in Figure
\ref{fig:stdardCase_snapshots}. The four snapshots
correspond to cycle minimum ($t_1$ = $t_{min}$, the first column), 
the time that the poloidal fields of a new cycle 
appear ($t_2$, $t_2$ = $t_1$ +  1.9 yr, the second column), 
cycle maximum ($t_3$, $t_3$ = $t_2$ +  2.1 yr = $t_{max}$, the third column), 
and the time that the latitudinal shear reaches a 
maximum ($t_4$, $t_4$ = $t_3$ +  4.0 yr, the fourth column).

Figures \ref{fig:stdardCase_snapshots}(a)-(d) show the
poloidal field evolution. About two years after $t_{min}$, the poloidal
fluxes of the new cycle appear (Figure \ref{fig:stdardCase_snapshots}(b)). The subsequent
transport and evolution of the newly generated flux
system have three branches. The first branch is the radially
downward transport, which is determined by the
radial pumping, radial component of the meridional flow, and turbulent diffusion. Above the depth of 0.88$R_\odot$,
the pumping dominates the transport process.
The pumping pushes the whole $\theta$-component of the
newly generated poloidal field below the depth of 0.89$R_\odot$ (see Figure \ref{fig:stdardCase_snapshots}(c)). The strong pumping along with the radial boundary condition causes the poloidal
field to be purely radial near the surface. This scenario
is consistent with that of \cite{Cameron2012}, who use
SFT models to constrain the BL-type dynamo at the surface.

%figure 6

In the depth range 0.7-0.89$R_\odot$, diffusion
and $u_r$ are responsible for the radial transport of the
poloidal field. 
% \textit{We define the advection timescale
% $\tau_U$ as the time taken for $u_r$ to
% advect the poloidal fields from $r=0.88R_\odot$ to $r=0.7R_\odot$.
% It is calculated by $\tau_U=L/u_{r0}$, where
% $L=0.18R_\odot$ is radial separation and $u_{r0}$ = 0.68 m s$^{-1}$
% is the average flow strength over 0.7-0.88$R_\odot$
% at $60^\circ$ latitude.
% The diffusion time across a length of $L$ is $\tau_D=L^2/4\eta_{CZ}$
% based on a Green's function for the diffusion equation
% \citep[for example, see][p.32]{Parker1979}.
% The values of $\tau_U$ and
% $\tau_D$ are 2.5 years and 6 years, respectively. Hence
% turbulent diffusion dominates the radial transport
% over the range 0.7-0.88$R_\odot$ with an almost
% homogenous velocity at different latitudes. Hence the
% $\theta$-component of the poloidal field, $B_{\theta}$,
% is dominant in the bulk of the convection zone. Thus we
% have the simple dipolar configuration of the poloidal
% field threading the two poles, consistent with the
% scenario suggested by \cite{Babcock1961}.} 
We have the simple dipolar configuration of the poloidal 
field threading the two poles, which is consistent with the scenario 
suggested by \cite{Babcock1961}. 
The latitudinal shear acting on $B_\theta$ immediately induces strong
toroidal fields in the whole convection zone (see Figures \ref{fig:stdardCase_snapshots}(j), (k), and (l)).
Figure \ref{fig:MC}(a) shows that in the range 
0.7-0.86$R_\odot$ the meridional flow is equatorward.
The equatorward flow is the reason for the equatorward migration of the toroidal
field that is presented in Figure \ref{fig:standard1_TorPol}(a).

The second branch of the surface poloidal field evolution is the poleward 
transport. This branch is advected by the poleward meridional flow 
and diffusion at the surface. The poloidal field contained in 
this branch cancels the existing polar field from the last cycle 
and then reverses the polarity of the polar field 
(see Figures \ref{fig:stdardCase_snapshots}(c) and (d)).
This explains the poleward migration of the surface
field shown in Figure \ref{fig:standard1_TorPol}(b) corresponding 
to the `magnetic butterfly diagram' \citep{Hathaway2015}.
The newly generated poloidal field
is simultaneously transported downward and poleward.
When $B_r$ reaches the polar regions, $B_\theta$ has been
transported downward to the bottom of the convection zone
(see Figures \ref{fig:stdardCase_snapshots}(b) and (c)). At $t_3$, the
source of the toroidal field $s(\textbf{B}_\theta\cdot\nabla_\theta\Omega)$  for an `old' cycle has
vanished and the toroidal field generation for a `new'
cycle starts (see Figure \ref{fig:stdardCase_snapshots}(k)). Therefore, when the polar
field reverses, the subsurface toroidal field reaches a
maximum at low latitudes (see Figure \ref{fig:stdardCase_snapshots}(g)).

As the cycle evolves, the newly generated poloidal flux
system moves closer to the equator (panels
($b$)-($d$) of Figure \ref{fig:stdardCase_snapshots}), 
which corresponds
to the third branch of the surface poloidal field evolution. The
equatorward migration pattern is due to the equatorward
migration of the subsurface toroidal field, which is the
source term of the poloidal field (panels ($b$)-($d$) and
($f$)-($h$) of Figure \ref{fig:stdardCase_snapshots}).
There are two causes of this equatorward migration of
the toroidal field: the classical cause
is the equatorward meridional flow
and the second cause is a new mechanism (as follows). Figure \ref{fig:diff_rot}(c) shows
that the latitudinal shear is the strongest around
60$^{\circ}$ latitudes. Its amplitude decreases at
higher and lower latitudes. Hence, the toroidal field of
a new cycle firstly appears around 60$^{\circ}$ latitudes.
It takes longer for the toroidal flux of a new cycle at
lower latitudes to be built up with the weaker
latitudinal shear.
This regeneration pattern of the toroidal field works like the equatorward
propagation. In addition, the across-equator coupling
of the poloidal field leads to the final solar-like
dipolar solution.

Figures \ref{fig:stdardCase_snapshots}(m)-(p) show the
contribution of the radial shear to the toroidal field generation.
The near-surface toroidal field generated by the radial
shear in the NSSL has high-latitude poleward and
low-latitude equatorward branches. The equatorward branch is
dominated by the Parker-Yoshimura rule for the direction
of dynamo waves \citep{Parker1955, Yoshimura1975}. 
The poleward branch results from the poleward migration of 
the surface $B_r$, which overpowers the dynamo wave at high 
latitudes. The toroidal field generated by the tachocline 
radial shear concentrates in the polar regions. This distribution 
provides strong and clear evidence that the tachocline 
is not the working place for our BL-type dynamo model. We further 
demonstrate the role of the tachocline in the following 
subsection.

% ** there is weak Br near the equator **

\subsection{A negligible role of the tachocline} \label{sec:negligibleRoleTachocline}
We have shown that in our dynamo model, the toroidal field
is mainly produced in the bulk of the convection zone by
the latitudinal shear. The radial shear of the tachocline
has a minor contribution to the toroidal field. Here, we
further demonstrate the negligible role played by the tachocline
in the toroidal field generation by a quantitative analysis
and two numerical tests.

\subsubsection{A quantitative analysis of the tachocline's effect}
\begin{figure}[t!]
  \gridline{\fig{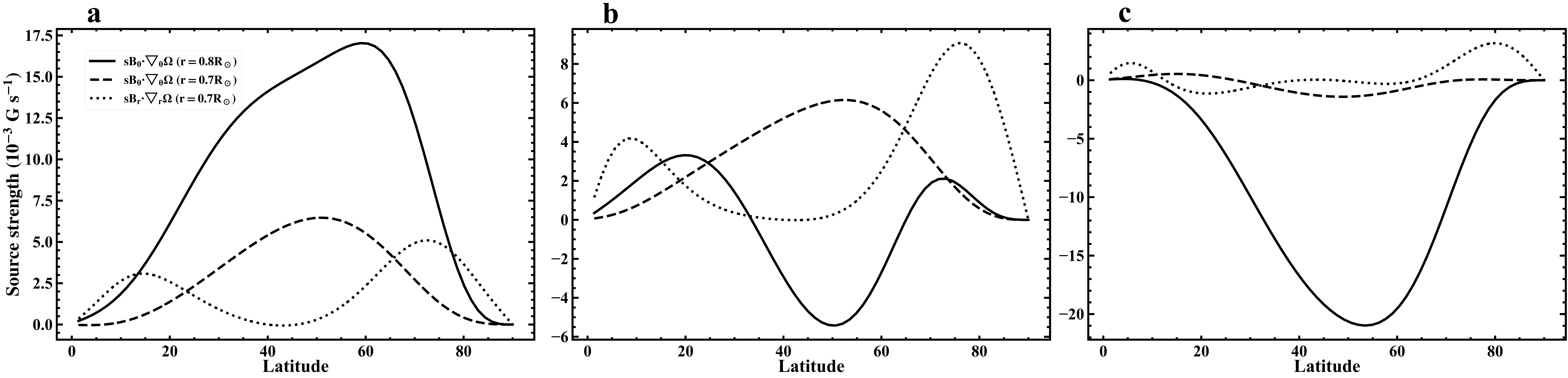}{1.0\textwidth}{}
            }
  \caption{Amplitude of the shear source terms as a
  function of latitude, at three times, $t_1$, $t_3$, and
  $t_4$, corresponding to panels (a)-(c) for the reference model. The solid
  curves represent the amplitude of
  $s(\textbf{B}_\theta\cdot \nabla _\theta\Omega)$ at
  middle of the convection zone, $r$ = 0.8$R_\odot$.
  The dashed and dotted curves represent the amplitude of
  $s(\textbf{B}_\theta\cdot \nabla _\theta\Omega)$ and
  $s(\textbf{B}_r\cdot \nabla _r\Omega)$
  at $r$ = 0.7$R_\odot$, respectively.}
  \label{fig:stdCase_source_lat}
\end{figure}

Now we compare the strength of the shear source terms
$s(\textbf{B}_{r}\cdot\nabla_r\Omega)$ and
$s(\textbf{B}_{\theta}\cdot\nabla_\theta\Omega)$ in
our reference model. Figure \ref{fig:stdCase_source_lat} shows results at three times,
$t_1$, $t_3$, and $t_4$, corresponding to the first, third,
and fourth columns of Figure \ref{fig:stdardCase_snapshots},
respectively. We consider the latitudinal dependence of $s(\textbf{B}_{r}\cdot\nabla_r\Omega)$ and
$s(\textbf{B}_{\theta}\cdot\nabla_\theta\Omega)$ at
\textbf{$r=0.8R_\odot$} and $r=0.7R_\odot$, which represent the
bulk of the convection zone and the tachocline, respectively.

%figure 7

At $t_{1}$, $s(\textbf{B}_{\theta}\cdot\nabla_\theta\Omega)$ at
$r=0.8R_\odot$ (Figure \ref{fig:stdCase_source_lat}(a), solid curve) 
peaks around 25$^\circ$ and 70$^\circ$. The average strength 
over latitudes $\langle s(\textbf{B}_{\theta}\cdot\nabla_\theta\Omega)\rangle$
is 8.8 G s$^{-1}$, which is about three times 
the value at $r=0.7R_\odot$ (dashed curve). The term 
$s(\textbf{B}_{r}\cdot\nabla_r\Omega)$ at
$r=0.7R_\odot$ (dotted curve) peaks around 80$^\circ$.
Its latitudinal averaged strength,
$\langle s(\textbf{B}_{r}\cdot\nabla_r\Omega)\rangle$,
is 1.9 $\times 10^{-3}$ G s$^{-1}$, which is about 
one fifth of $\langle s(\textbf{B}_{\theta}\cdot\nabla_\theta\Omega)\rangle$
at $r=0.8R_\odot$. If we exclude the values above
50$^\circ$, $\langle s(\textbf{B}_{r}\cdot\nabla_r\Omega)\rangle$
is close to zero. 
% \textbf{At $t_1$, the latitudinal shear in the bulk of the 
% convection zone is the dominated generator for the toroidal flux.
% And its strength is decaying since the poloidal field created
% at the surface for the new cycle is transported downward. 
% These poloidal fields cancel the poloidal field of the old 
% cycle. At $t_{2}$, the sign of the latitudinal shear at the
% middle of the convection zone changes, which means that
% the dominated generator of the toroidal flux for the old 
% cycle stops. Therefore, the moment $t_2$ corresponds to 
% the cycle maximum naturally.}
At $t_{1}$, the shear strength
is decaying because the poloidal field is being canceled.
About two years from $t_1$, the poloidal fluxes of
the new cycle appear at the surface
(Figure \ref{fig:stdardCase_snapshots}(b)). 
It takes about two years to transport these fluxes to the
poles and the bottom of the convection zone. Therefore,
at $t_3$ (Figure \ref{fig:stdCase_source_lat}(b)) the
sign of $s(\textbf{B}_{\theta}\cdot\nabla_\theta\Omega)$
changes and the toroidal field of the new cycle, whose
polarity is opposite to that of the previous cycle,
begins producing at middle latitudes
(see Figure \ref{fig:stdardCase_snapshots}(g)). The
distribution of $s(\textbf{B}_{\theta}\cdot\nabla_\theta\Omega)$
at $r=0.8R_\odot$ peaks around 50$^\circ$. 
%\textit{ At $r=0.7R_\odot$,
% $s(\textbf{B}_{\theta}\cdot\nabla_\theta\Omega)$
% and $s(\textbf{B}_{r}\cdot\nabla_r\Omega)$ become very weak since the
% newly generated poloidal flux just reaches $r=0.7R_\odot$ and
% canceles a large part of the poloidal flux
% of the last cycle.} 
With the continuous generation and downward diffusion of the
newly generated poloidal fields, the strength of
$s(\textbf{B}_{\theta}\cdot\nabla_\theta\Omega)$ reaches
a maximum at $t_4$ (Figure \ref{fig:stdCase_source_lat}(c)).
At this moment, the average strength of
$s(\textbf{B}_{\theta}\cdot\nabla_\theta\Omega)$ at
$r=0.8R_\odot$ is 9$\times 10^{-3}$ G s$^{-1}$, 
which is about 30 times the value at $r=0.7R_\odot$. The average
strength of the radial shear is 0.4 $\times 10^{-3}$
G s$^{-1}$ at $r=0.7R_\odot$, which is one twentieth of 
the average value of $s(\textbf{B}_{\theta}\cdot\nabla_\theta\Omega)$
at $r=0.8R_\odot$. If we exclude the values above 50$^\circ$,
$\langle s(\textbf{B}_{r}\cdot\nabla_r\Omega)\rangle$ at
$r=0.7R_\odot$ is also close to zero.

Based on the quantified analysis of the shear strength
at the three moments, we confirm that most toroidal
fluxes come from the latitudinal shear in the bulk of the convection
zone. 
The radial and latitudinal shear in the tachocline 
have a weak contribution to the toroidal field. 

\subsubsection{Numerical experiments}
In the reference model, the differential rotation profile is 
based on helioseismic results, including the tachocline 
characterized by the strong radial shear at the base of the 
convection zone. Since the tachocline has been shown
to play a minor role 
in our dynamo model, we now carry out numerical experiments to 
check the dynamo behavior by artificially shutting off the 
tachocline in two ways.

% %figure 8
\begin{figure*}[t!]
  \gridline{\fig{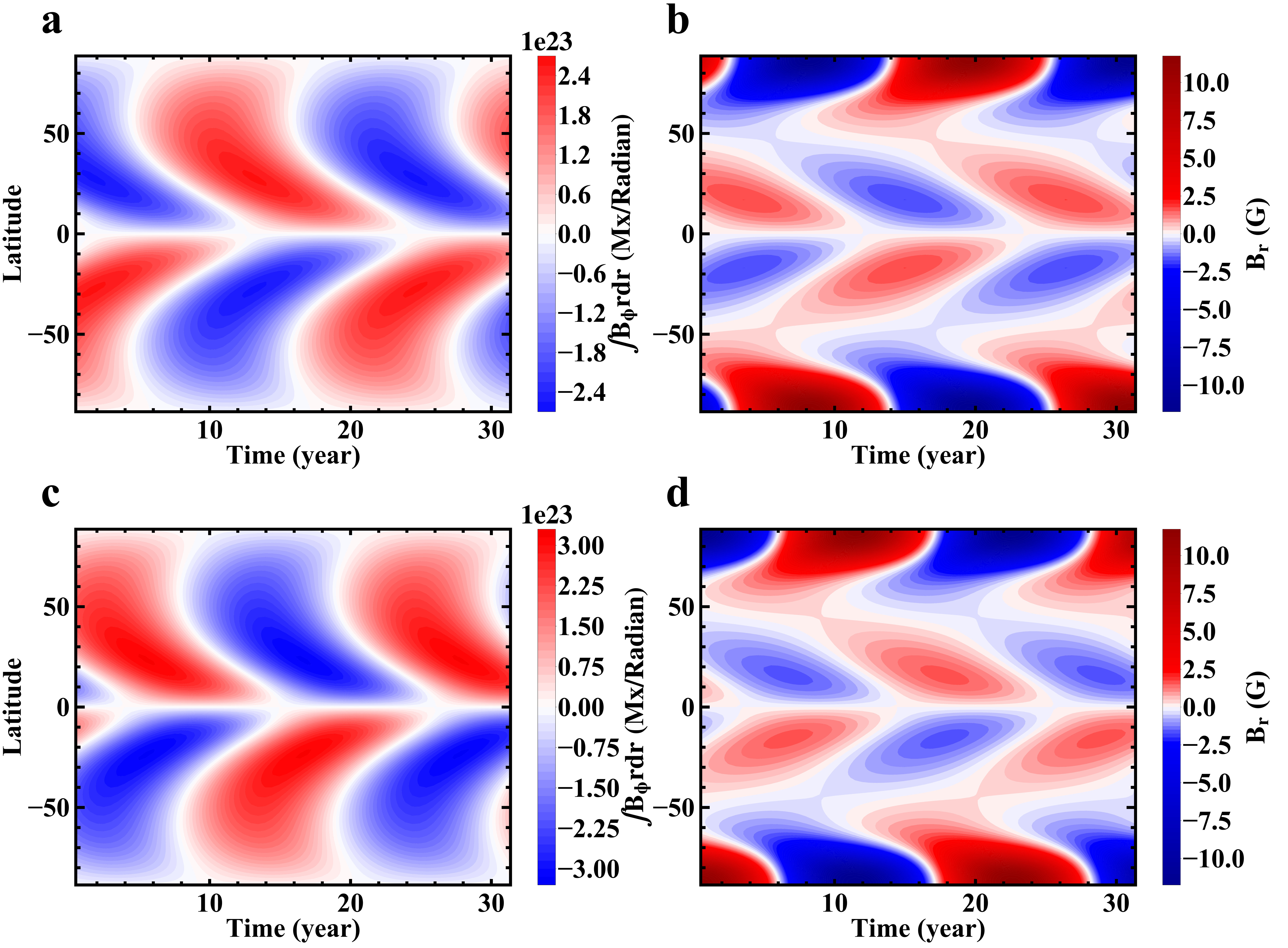}{0.8\textwidth}{}
            }
  \caption{Same as Figure \ref{fig:standard1_TorPol}, 
  but panels (a) and (b) for the case with the differential 
  rotation shown in Figure \ref{fig:diff_rot}(b); panels
  (c) and (d) for the case with the bottom boundary at 
  0.725$R_\odot$.}
  \label{fig:resultsNoTac}
\end{figure*}

In the first experiment, we use the differential rotation
profile shown in Figure \ref{fig:diff_rot}(b) to artificially remove the
radial shear in the tachocline. The latitudinal shear of
the tachocline is still included in the experiment, but it is 
slightly stronger than that of the original profile.
We keep the other ingredients of the model the same as those in 
the reference model, including other model parameters, initial and 
boundary conditions, except $\alpha_0$ and $\eta_{CZ}$.
Since the average strength of the latitudinal shear in the 
first experiment is stronger than that in the reference model,
as shown in Figures \ref{fig:diff_rot}(c) and (d),
to keep $\nabla \Omega R_\odot^2 /\eta$ in this 
experiment, consistent with that in the reference model, 
we use a larger diffusivity, 
$\eta_{CZ}=2.9\times10^{11}$ cm$^{2}$ s$^{-1}$, than that 
in the reference model. Accordingly, a 10$\%$ supercritical 
$\alpha_0$ = 0.92 m s$^{-1}$ is used. The results are 
shown in Figures \ref{fig:resultsNoTac} (a) and (b),
which are derived in the same way as those in 
Figure \ref{fig:standard1_TorPol}.
% Figure \ref{fig:resultsNoTac} gives the time evolution of
% the radially integrated toroidal field and surface radial
% field, which are derived in the same way as that in 
% Figures \ref{fig:standard1_TorPol}.
The solution has a 11.4 yr cycle period and a similar field 
distribution as that of the reference model. 
% But both the toroidal and
% poloidal fields are stronger than that of the
% reference model. The differential rotation profile
% used here has a bigger value of
% $(1/r)(\partial \Omega /\partial \theta)$ at the bottom of the
% convection zone than that used in the reference model
% (see Figures \ref{fig:diff_rot}(c) and (d)). The increase of the
% latitudinal shear causes the increase of both the poloidal
% and toroidal fields. The ratio between the poloidal
% and toroidal fields is similar to the reference model.

In the second experiment, we set the bottom boundary at
a depth of 0.725$R_\odot$, which means that the effects
of the tachocline on the dynamo model are completely
removed. The penetration depth of 
the meridional flow is changed to $r_b$ = 0.725$R_\odot$. 
The parameter $\psi_0/C$ is changed to 39.1 m s$^{-1}$ to 
ensure a maximum surface flow of 20 m s$^{-1}$. 
We use 
$\eta_{CZ}=3.1\times10^{11}$ cm$^{2}$ s$^{-1}$ and
$\alpha_0$ = 0.63 m s$^{-1}$ in this case, for the same 
reason shown in the first experiment.
The other parameters are the same as those in the
reference model. The results are shown in Figures \ref{fig:resultsNoTac} (c) and (d).
% that the meridional flow
% parameter $r_b$ is changed from 0.65$R_\odot$ to 0.725$R_\odot$.
The dynamo could also operate successfully for this case.
Both the toroidal and poloidal fields show a similar pattern as 
those in the reference model, and the cycle period is 11.1 yr.  
In summary, the two numerical tests further confirm the 
negligible role played by the tachocline in our model.
% , except that the cycle
% period is 0.5 yrs shorter. The strength of the toroidal
% and poloidal fields is about 10 times stronger than
% that of the reference model. Their strength can be
% decreased by the decrease of the $\alpha_0$ value.
% We do not show the results here.

\subsection{Comparison with Other representative models} \label{sec:ComparisonsModels}
\begin{figure}[ht!]
  \gridline{\fig{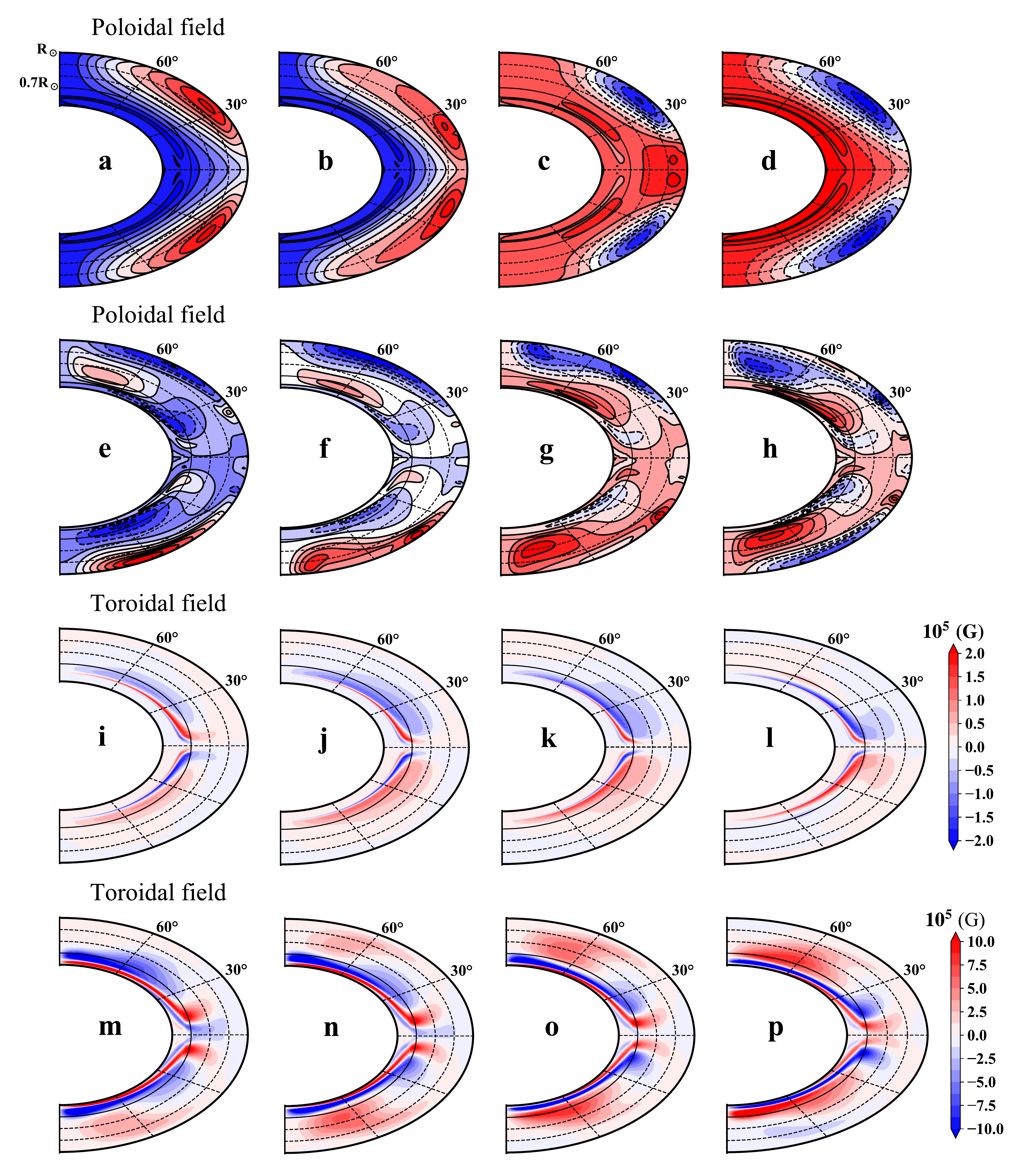}{1.0\textwidth}{}
            }
  \caption{Snapshots of the poloidal field (first and second rows) and
  toroidal field (third and fourth rows) for
  model CNC2004 (first and third rows)
  and model DC1999 (second and fourth rows) over one half 
  cycle for comparison with the reference model shown in 
  Figure \ref{fig:stdardCase_snapshots}. Each column is 
  advanced by an eighth of the dynamo cycle;
  that is, from left-hand to right-hand $t$ = 0, $T$/8, $T$/4, and 3$T$/8.
  Solid (dashed) poloidal field lines are clockwise (counterclockwise).}
  \label{fig:modelsComp}
\end{figure}

\cite{Dikpati1999}, hereafter DC1999, and
\cite{Chatterjee2004}, hereafter CNC2004, are two
representative FTD models. \cite{Yeates2008} gave a close
investigation of the two models. Here, we compare our
reference model, hereafter ZJ2021, with them to illustrate
the differences among the models. To reproduce DC1999, we take
the set of parameters listed in their Section 3. The only
difference is that we extend the solution to both the
northern and southern hemispheres rather than only the
northern hemisphere, as used in their model. For CNC2004, we use
the parameters in their Section 3.1. Figure \ref{fig:modelsComp} shows the time
evolution of the poloidal and toroidal fields in the two
models for four successive intervals during a half of the
dynamo period.

% %figure 9

A prominent difference among the three models is the
configuration of the poloidal field. In our model, the
poloidal field is highly organized. The near-surface
field is purely radial. At cycle minimum, the poloidal
field has a global and simple dipolar structure. Panels
($a$)-($d$) of Figure \ref{fig:modelsComp} show the poloidal field evolution
from CNC2004. Although there is also a cross-equator
connection of the poloidal field that is similar to ours,
dissipation of the poloidal field crossing the surface
exists. Circular structures of the poloidal field near
the surface are an implication of the dissipation and
correspond to a strong $B_\theta$ component, which is
not consistent with the prerequisite of the empirical
SFT models \citep{Cameron2012}. Panels ($e$)-($h$) of Figure \ref{fig:modelsComp} show the poloidal
field evolution from DC1999. The poloidal field has a
complex configuration. In contrast to the global structure in
our model, several local structures in both latitudinal
and radial directions are presented here. Moreover, the
poloidal fields in two hemispheres have an
antisymmetric structure, which corresponds to
a quadrupolar solution.

Besides the different outer boundary conditions (vertical versus 
potential), the different flux transport mechanisms of the 
poloidal field are mainly responsible for the different 
configurations of the poloidal field. In our model, the downward 
pumping along with the radial outer boundary condition keeps the
poloidal field radial near the surface. The strong
turbulent diffusion dominates the poloidal field
transport through the bulk of the convection zone. Since
the polar field is radial, the meridional flow, which is also radial near the poles,
cannot subduct the polar field into the interior. However, in both CNC2004 and
DC1999, there is a subduction of the poloidal field by the
meridional flow sinking underneath the surface at the
polar regions. The subducted flux plays a significant role in their magnetic
field evolution \citep{Hazra2017}. Especially in DC1999,
the meridional flow dominates the transport of the
poloidal field. The surface polar field dominated by the
$B_\theta$ component is advected downward to the
tachocline. Figure \ref{fig:modelsComp}(h) shows the
highly concentrated polar field, which causes
a low ratio of $B_{\phi\max}/B_{r\max}$. The strong polar
field and weak ratio of $B_{\phi\max}/B_{r\max}$ have been 
regarded as difficulties of the model. In CNC2004, a
part of the newly generated poloidal fields is transported
downward by diffusion. Compared with our model, the downward
$B_\theta$ is relatively weak because a part of the poloidal
fields is dissipated across the surface.
All of these ingredients cause the poloidal field
evolution, including the polar field evolution, among the
three models to be strikingly different.
% All the
% above-mentioned ingredients cause that the poloidal field
% evolution including the polar field evolution among the three models is
% strikingly different.

Different configurations of the poloidal field further
determine where the toroidal field is produced. In our 
model, the poloidal field has a global scale and is 
$\theta$-component dominated in the bulk of the 
convection zone. This poloidal field configuration 
determines that most toroidal fields are produced
by the latitudinal shear in the bulk of the convection 
zone. The latitudinal shear peaks around middle 
latitudes. Along with the effect of the equatorward 
meridional flow, most of the toroidal fields are 
produced at mid-low latitudes.  There are no strong 
toroidal fields at the polar regions.
In CNC2004, there is a quite strong $B_\theta$ 
distribution in the bulk of the convection zone, but $B_\theta$ in
the tachocline is even stronger. There is a certain amount of
$B_r$ in the tachocline. Hence, the newly generated
toroidal field concentrates at the bottom of the
convection zone (see Figures \ref{fig:modelsComp}(i)-(l)). 
Both the latitudinal and radial shears of the tachocline 
contribute to the toroidal field generation.
\cite{Jaramillo2009} have pointed out that
the latitudinal shear is dominated in
producing the toroidal field in models with a high
diffusivity. In DC1999, effects of the latitudinal shear
on the toroidal field generation were also addressed.
Since the poloidal field is dominated
by the $r$-component in the tachocline, the radial
shear of the tachocline contributes most to the toroidal
field generation (see Figures \ref{fig:modelsComp}(m)-(p)). The radial shear of the tachocline peaks at high
latitudes. The concentrated toroidal field at high
latitudes is inevitable. Note that in both CNC2004 and 
DC1999, the toroidal field concentrates below 0.75$R_\odot$ 
(see panels (i)-(p) of Figure \ref{fig:modelsComp}), which 
is strikingly different from our reference model.

%figure 10
\begin{figure}[t!]
  \gridline{\fig{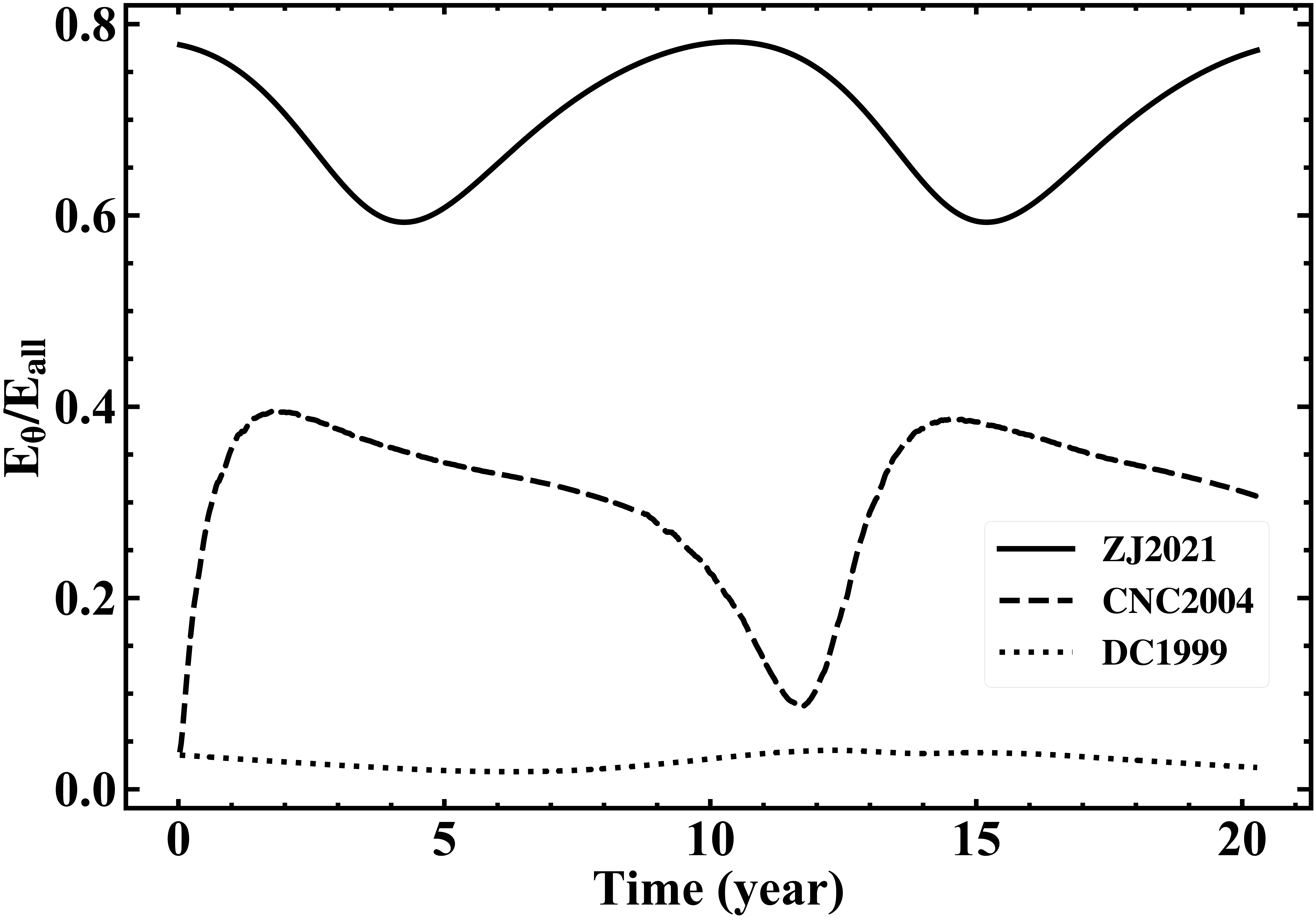}{0.5\textwidth}{}
  }
  \caption{Comparison of the time evolution of the magnetic 
  energy ratio, $\frac{E_\theta}{E_{all}}$ defined by Eq.(\ref{eq:energyRatio1}) 
  among ZJ2021 (solid line), CNC2004 (dashed line), and DC1999 (dotted line).}
  \label{fig:energyRatio}
\end{figure}

We now give a quantitative comparison of the relative strength of $B_\theta$ among 
the three models. Figure \ref{fig:energyRatio} shows the time evolution of the magnetic energy ratio of
\begin{equation}
  \frac{E_\theta}{E_{all}} = \frac{\int_{0.7R_\odot}^{0.89R_\odot} \int_{30^\circ}^{90^\circ}B_\theta^2r^2\sin\theta  \,dr \, d\theta}
  {\int_{0.7R_\odot}^{R_\odot} \int_{0^\circ}^{90^\circ}(B_\theta^2+B_r^2)r^2\sin\theta  \,dr \, d\theta }
  .\label{eq:energyRatio1}
\end{equation}
The maximum value of $E_\theta/E_{all}$ in our reference model is
0.8, which is about twice as large as that of CNC2004 and about 
20 times larger than that of DC1999.
During the whole period, $B_\theta$ dominates the 
poloidal field in our model. 
% \textbf{This high value reflects the inhibition effect of the 
% pumping for the outward diffusion of the poloidal field. 
% Most of the poloidal field are constrained within the 
% convection zone. In our model, $B_\theta$ 
% dominates the poloidal field, corresponding to the global 
% dipolar field. And they are wound up by the latitudinal 
% differential rotation.} 
% \textit{Figure \ref{fig:energyRatio}(b) shows the time evolution of the magnetic energy ratio of
% \begin{equation}
%   \frac{E_\theta}{E_r} = \int_{0.7R_\odot}^{0.88R_\odot} \int_{30^\circ}^{90^\circ}B_\theta^2/B_r^2 \,dr \, d\theta .\label{eq:energyRatio2}
% \end{equation}
% During the whole period, $B_\theta$ dominates the 
% poloidal field in our model. The maximum value of $E_\theta/E_r$
% is about 223, which is about 6 times larger than that of
% CNC2004 and about 50 times larger than that of DC1999.}
% To further illustrate the difference in the radial location of the
% toroidal field among the three models, we present the time evolution
% of $E_{tor}$ from CNC2004 and DC1999 in Figure \ref{fig:CNC_DC_T_r}.

In summary, our model has a prominently different
configuration of the poloidal field from that of
DC1999 and CNC2004. Different flux transport
mechanisms and outer boundary conditions account
for the different poloidal field configurations,
which lead to the different locations of the
toroidal field generation.

\section{Conclusion} \label{sec:con}
We have developed a new BL-type solar dynamo model,
in which the toroidal field is generated in the bulk of the
convection zone by the latitudinal differential rotation.
The radial shear in the tachocline plays a negligible role
in the toroidal field generation. This distinguishes 
our model from the popular view
that the tachocline is the location of the toroidal magnetic flux production, 
yet fulfills the needs of recent advances in the 
understanding of solar magnetism (as listed in our introduction).
The model satisfactorily reproduces the following
major solar cycle features: (1) about 11 yr cycle period
and 18 yr extended cycle period, (2) the equatorward
migration of the toroidal fields and the poleward drift
of the large-scale surface radial fields, (3) the
realistic phase relation between the polar field and
the toroidal field, and (4) a solar-like antisymmetric
magnetic field about the equator. We have also carried 
out two numerical experiments to check the dynamo
behavior by artificially shutting off the tachocline. 
These two experiments show similar solutions to the reference model.

Our model has several advantages as compared with DC1999
and CNC2004 due to the following facts. First, the strong 
polar branch of the toroidal field is circumvented in 
our model. 
The toroidal flux of each cycle starts from the polar 
region, with weak strength corresponding
to ephemeral regions. With the equatorward migration 
of the toroidal flux, its strength increases and
corresponds to active regions at middle and low latitudes. 
Thus, the results account for not only the
solar cycle but also the extended cycle. In contrast, 
the strong polar branch is a typical property of the 
FTD models working in the tachocline. To remove the 
branch, additional assumptions have to be included, 
such as the deep penetration of the meridional
flow \citep{Nandy2002, Chatterjee2004}.

Second, we obtain a solar-like antisymmetric magnetic 
field about the equator with the pure BL-type $\alpha$-effect 
alone in our model. An extra $\alpha$-effect is required 
in FTD models to account for the dipolar 
solution \cite[e.g.,][]{Dikpati2004, Passos2014}.
Third, the toroidal and poloidal fields have the
same strong diffusivity, which is
2$\times10^{11}~$cm$^2$s$^{-1}$ in the bulk of the 
convection zone. It is still
weaker than the estimated value based on the MLT but
is one order higher than the value used in previous
FTD models. The quenching of the turbulent
diffusivity for the toroidal field, which is adopted by
CNC2004, is not plausible based on \cite{Karak2014}.

The most significant advantage of our model lies in
producing a realistic configuration and evolution of
the polar field. The polar field peaks around the cycle
minimum, with an amplitude of about 10 G. It is
radial, hence there is no subduction of the polar
flux by the radially downward meridional flow sinking
underneath the surface. Thus, the evolution of the polar
field is purely due to the flux transport from the
activity belt. In contrast, previous FTD models tend to 
generate a polar field that is too strong \citep[e.g.,][]{Durney1997, Dikpati1999, Hazra2017}.
The meridional flow subducts the polar field sinking underneath 
the surface. The polar flux submergence
plays an important role in their polar field evolution.

One of the reasons for the effectiveness of our model is 
that it includes near-surface radial pumping and a radial 
outer boundary condition. These
ingredients are derived from the constraints of
surface flux evolution by \cite{Cameron2012}
and they lead to a simple dipolar configuration
of the poloidal field, which contains the purely
radial field near the surface and the latitudinal field within the convection
zone. Consequently, the latitudinal shear winds up the
global dipole to produce most of the toroidal
fields in the bulk of the convection zone. This
model is consistent with Babcock's original
scenario. The BL-type solar dynamo models developed
by \cite{Kitchatinov2012} and \cite{Kitchatinov2017b} also do
not depend on the tachocline. In addition, the diamagnetic pumping
is included in their models, but the downward
pumping is very efficient near the bottom of the
convection zone. As a result, although the toroidal field is not generated in
the tachocline, it is near the
base of the convection zone ($\sim 0.72 - 0.81 R_\odot$), where the diffusivity
has a rapid variation over 3 orders, from lower than $10^{9}$ cm$^2$ s$^{-1}$
to higher than $10^{12}$ cm$^2$ s$^{-1}$. Furthermore, 
because the radial field
assumption is not satisfied, the evolution of the surface field cannot be
guaranteed to be consistent with that of the SFT models.

The flux transport mechanisms are another 
reason for our model's success.
The generating layers of toroidal and poloidal fields are spatially segregated in FTD models.
Toroidal and poloidal fields are connected by flux transport mechanisms, especially the
meridional flow. The meridional flow plays an essential role in
regulating the cycle period and the equatorward
migration of the toroidal field in FTD models. In contrast, the generation layers of toroidal and poloidal fields
are partially coupled in our model, due to the effect of surface radial pumping.
Furthermore, the meridional flow cannot transport the polar 
flux sinking under the surface. Hence, our model is less reliant on the inner structure
of the meridional flow than FTD models. In a forthcoming
paper, we will explore this property in detail.




\begin{acknowledgments}
  This research was supported by the National Natural Science Foundation of China through grant Nos. 11873023 and 12173005, the Fundamental Research Funds for the Central Universities of China, Key Research Program of Frontier Sciences of CAS through grant No. ZDBS-LY-SLH013, the B-type Strategic Priority Program of CAS through grant No. XDB41000000, the International Space Science Institute Teams 474 and 475.
  \end{acknowledgments}

\bibliography{sample631}{}
\bibliographystyle{aasjournal}



\end{document}